# Analysis of the ordering transition of hard disks through the Mayer cluster expansion

E. Eisenberg[1] and A. Baram[2]

1. School of Physics and Astronomy, Raymond and Beverly Sackler Faculty of Exact Sciences, Tel Aviv University, Tel Aviv 69978, Israel

2. Soreq NRC, Yavne 81800, Israel

**Abstract**. The available virial coefficients for the two-dimensional hard–disks model are transformed into a matrix representation of the thermodynamic potentials, which allows for an accurate description of the whole fluid phase, up to the phase transition. We find that the fluid phase terminates at the transition point, supporting the Kosterlitz-Thouless-Halperin-Nelson-Young picture of a second order phase transition into a Hexatic phase. The density and pressure at the transition are calculated from the available first ten virial coefficients, and are found to be in excellent agreement with recent Monte-Carlo calculations. Finally, we calculate the equation of state in the critical region.

It has become clear recently that the essential molecular mechanism that drives freezing transitions can be understood in terms of entropy. This is demonstrated by ordering transitions observed in purely entropy-driven hard-core models, such as the hard spheres model which has played a key role in a statistical description of freezing. The equation of state of classical hard-spheres in d dimensions is a long-standing problem of great importance in statistical mechanics. Since the pioneering works of van der Waals [1] and Boltzmann [2], these systems have been investigated intensively, but there are still a lot of fundamental unresolved problems. In particular, the analytic properties of the freezing transitions of the fluid phases (for d>1) are unknown. These transitions have been demonstrated so far only by Monte-Carlo (MC) computer simulations [3-6]. For d≥3 the transition is a first order phase transition, while for the hard disks system (d=2) it is believed to be a second order transition from the fluid phase to a orientationally ordered hexatic phase [5,6], in accordance with the Kosterlitz-Thouless-Halperin-Nelson-Young (KTHNY) picture[7-9]. There is a huge gap between the values of the MC freezing densities: $\rho_f(d=2) = 0.779(2)$ [5,6] and $\rho_f(d=3) = 0.663(2)$ [10,11] (densities here are in units of the corresponding densities of closest packing $\rho_{cp}(d)$), and the best known rigorous lower bounds for the termination densities of these fluid phases [12]: $\rho_{lb}(d=2) = 0.2849$ and $\rho_{lb}(d=3) = 0.17125$.

Many approximate equations of state have been proposed, most of them based on the available virial coefficients. However, none of these proposed equations exhibits a freezing transition at the above MC densities. Indeed, for a variety of hard core lattice-gas models the asymptotic behavior of the virial coefficients is determined by a pair of non physical complex singularities in the complex $\rho$ plane that mask the physical singularity on the real axis [13-16]. As a result, it is very difficult to extract information concerning the termination of the fluid phase or the freezing transition from the virial series.

In this work we apply a transformation of the virial series into a matrix representation of the thermodynamic functions as a function of the activity. It is shown that this transformation results in a tractable asymptotic behavior of the matrix elements for repulsive systems. It thus allows for extrapolation of the existing virial coefficients to yield an accurate description of the fluid phase up to its termination point. In particular, we show the hard-disks fluid terminates at the ordering transition point, showing no evidence for a super-cooled disordered phase or random closest packing, in contrast with the hard-spheres scenario. This supports the KTHNY picture, suggesting a second order phase transition at the termination of the fluid branch. We accurately predict the density $\rho_c = 0.795(10)$ and pressure $p_c=7.95(2)$ at the phase transition in agreement with recent MC results, and find the critical dependence of the equation of state (the pressure as a function of density) in the vicinity of the transition.

The low density Mayer cluster expansion, in terms of the activity z, provides an alternative description of the fluid phase. In contrast to the virial series, the asymptotic form of the cluster coefficients of a repulsive system is well defined [17-19]:

$$nb_n \cong (-z_0)^{-(n-1)} n^{-\phi(d)} \{1 + cn^{-\theta(d)}\} \tag{1}$$

The radius of convergence of the cluster series, $z_0$, is model-dependent, but the exponents $\phi(d)$ and $\theta(d)$ are universal, depending only on the spatial dimensionality d. It turns out that a convenient representation of the equation of state can be obtained in terms of the real symmetric tri-diagonal matrix, R, defined by:

$$(R^n)_{11} = (-1)^n (n+1) b_{n+1}. \tag{2}$$

The first elements of the R matrix are given by:

$$\begin{aligned} R_{11} &= -2b_2 \\ R_{12}^{\,2} &= 3b_3 - 4b_2^2 \\ R_{22} &= [-4b_4 + 12b_2 b_3 - 8b_2^3]/[3b_3 - 4b_2^2] \end{aligned} \tag{3}$$

The density is then given in terms of the R matrix by:

$$\rho(z) = \sum_{n=1}^{\infty} n b_n z^n = \sum_{n=0}^{\infty} (-1)^n z^{n+1} \left(R^n\right)_{11} = z(I + zR)^{-1}{}_{11}, \qquad (4)$$

where I is the identity matrix. These matrix elements can be represented by the shifted Stieltjes summation [20]:

$$\rho(z) = \sum_{j=1}^{\infty} u_j \frac{z}{1 + \lambda_j z} \qquad (5)$$

where the eigenvalues of R, $\lambda_j$, are just the Yang-Lee zeroes of the grand canonical partition function, and $u_j$ is the square of the first component of the j'th eigenvector of R and thus satisfies $u_j > 0$ for all j. The spectrum lies in the range $-z_t^{-1} \leq \lambda_j \leq z_0^{-1}$, where $z_t$ is the termination activity of the fluid phase. As a result, the matrix representation has a tractable form in spite of the fact that the ratio $z_t/z_0$ is large. We have recently applied this matrix approach to the study of the hard core $N_3$ square lattice model, and provided evidence concerning the critical region and the nature of the fluid-solid transition [21]. In particular we showed that the fluid phase does not terminate at $z_c$, the critical activity, where the system solidify through a first order phase transition.

Recently, Clisby and McCoy [22] have evaluated the virial coefficients $B_n(d)$ for hard spheres in d dimensions up to n=10 and d=8. In this letter we use their results to obtain an accurate equation of state for the hard disks system (d=2), employing the above matrix representation.

Our first step is to estimate the location of the leading singularity $z_0$. Utilizing the ten available virial coefficients and the fact that $\phi(2) = \theta(2) = 5/6$, we applied the Levin acceleration method [23] to the series $\{n^{11/6}|b_n|\}$. Inspecting the location of the divergence of the various approximants, we estimated the value $z_0 = 0.168476(1)$ for the leading singularity.

We then calculated the first nine non-vanishing matrix elements of the R matrix (table 1). As expected, the diagonal and near diagonal matrix elements rapidly converge to constant asymptotic values, forming a Toeplitz-like matrix whose asymptotic constant values B (diagonal) and A (near diagonal) are related to the two branch points of the fluid thermodynamic functions by

$$z_0^{-1} = 2A + B \qquad z_t^{-1} = 2A - B \qquad (6)$$

Table 1: First nine elements of the R matrix. The numbers in parenthesis are the errors in the least significant digit (one standard deviation), resulting from the errors in the coefficients $B_n$ as reported in Ref. [22]

| n | 1 | 2 | 3 | 4 | 5 |
|---|---|---|---|---|---|
| $R_{nn}$ | 3.6275987 | 2.9629630 | 2.965658(5) | 2.966474(53) | 2.96718(29) |
| $R_{n,n+1}$ | 1.6494542 | 1.5240013 | 1.501923(14) | 1.494173(68) | |

Figure 1 presents the diagonal matrix elements $R_{nn}$ and twice the near diagonal terms $2R_{n,n+1}$ as a function of $1/n^2$, showing that the matrix elements converge towards the asymptote like $1/n^2$, consistently with the lattice N3-model behavior [21]. The intercepts of the resulting straight lines are: B=2.9675(3) and 2A=2.9680(3), in perfect agreement with the previous independent estimate $z_0^{-1} = 2A + B = 5.93556(4)$. These extrapolated values lead to $z_t^{-1} = 2A - B = 0.0005(5)$, which yields the bound $z_t \geq 1000$ for the termination activity of the fluid phase.

While we are not able to locate the termination activity accurately, it turns out that the termination density is not sensitive to the exact value of $z_t$. For fixed asymptotic values A and B, or equivalently fixed values of the two branch points $z_0$ and $z_t$, we extrapolate the matrix elements of the R matrix assuming $n^{-2}$ convergence towards the asymptotic values, and then compute the density $\rho(z)$. We find that the density is insensitive to the small uncertainties in the value of the leading singularity $z_0$, and is independent of $z_t$ for $z \ll z_t$ (figure 2). Furthermore, the value of the termination density $\rho(z_t) = 0.795(10)$ is almost independent on $z_t$ for all values of $z_t$ obeying $z_t \geq 1000$. This value of the termination density is indeed very close to the MC fluid–hexatic transition density $\rho_c = 0.779(2)$ [5,6]. It is tempting to conjecture that both values are identical and the fluid phase does indeed terminate at the activity $z_t = z_c$. Termination of the fluid phase at the transition point is characteristic of a second order phase transition, in contrast with the situation in first order fluid-solid phase transitions (such as the hard-spheres and the N3 [21] solidification transitions), where the fluid branch can be continued beyond the transition, describing a metastable super-cooled fluid. Thus, our results support the KTHNY scenario of the hard disks fluid terminating at a second order phase transition towards the haxatic phase [5-9].

Unlike the density, the pressure at the transition is sensitive to $z_c$. Thus, we are not able to predict the transition pressure based on the virial coefficients alone. Instead, we adopt the value suggested in the literature for the termination activity, $z_t=z_c\sim\exp(12.7)$ [24], and calculate the pressure based on the above matrix method given this value (figure 3). This calculation provides us with the full equation of state, covering the whole fluid branch from zero density towards the transition. In particular, it predicts $p(z_c)=7.95(2)$, in excellent agreement with recent MC calculations [6], further attesting towards the accuracy of our approach.

Finally, we study the critical behavior in the vicinity of the transition. Regardless of $z_t$, the density in the region $z \approx z_t$ is well fitted by:

$$\rho(z) = \rho_c - c(z_c - z)^{1-\alpha} \tag{7}$$

and the critical exponent is (independent of $z_t$), $\alpha \cong 0.75(3)$. Accordingly, the pressure has a strong dependence on the density in the vicinity of the transition, the leading term of which is given by (see inset of figure 3):

$$p_c - p \approx (\rho_c - \rho)^{1/(1-\alpha)} \tag{8}$$

This strong density dependence might account for the difficulties experienced by many groups using various approaches in probing the transition region through the virial expansion.

The extrapolation method through the R matrix as presented here is applicable, in principle, to all repulsive potential systems. In particular, it could be used to study hard spheres systems in any spatial dimensionality. In practice, it depends on the accuracy of the available virial coefficients. Determination of the rate of convergence of the R matrix elements to their asymptotic values is sensitive to uncertainties in the values of the matrix elements, as well as to the corrections to the asymptotic behavior of these matrix elements.

In summary, it is shown that the available Mayer cluster expansion coefficients, transformed into a matrix representation, allow for a reliable description of the hard disk fluid properties, up to the its termination point. The fluid phase terminates at the transition point, as expected for a second order phase transition, with no metastable super-cooled fluid phase. We also find the behavior of the equation of state in the critical region. These results are in concordance with the KTHNY picture.

**Figures:**

**Figure 1**: The diagonal $R_{nn}$ (circles) and twice the off-diagonal $2R_{n,n+1}$ (triangles) elements of the tri-diagonal matrix R as a function of n, and their extrapolation for n>>1, as a function of $1/n^2$ (all error bars are smaller than symbol sizes).

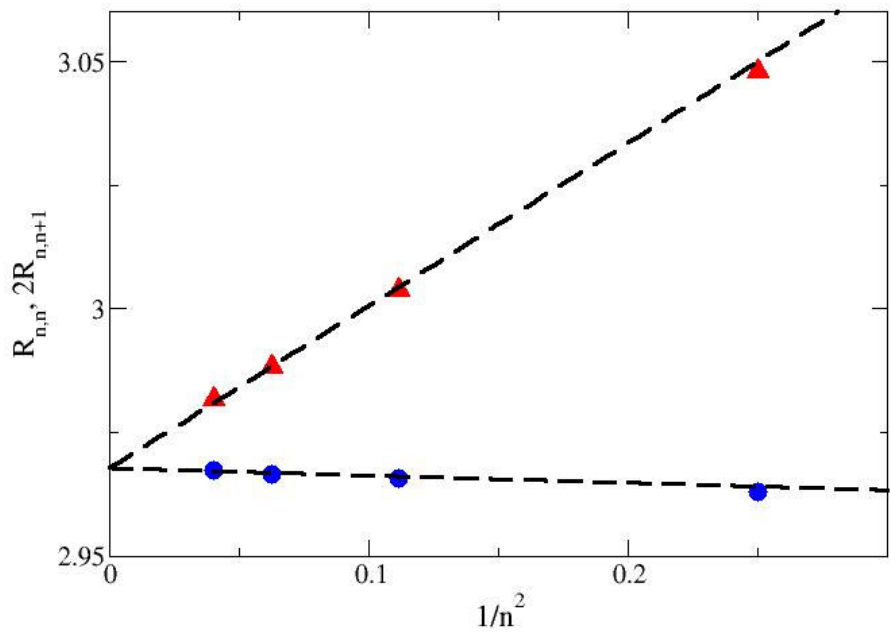

**Figure 2**: The density as a function of the activity for various choices of the termination activity $z_t$. The low density behavior is independent on $z_t$. Moreover, the termination density $\rho(z_t)$ depends only weakly on $z_t$.

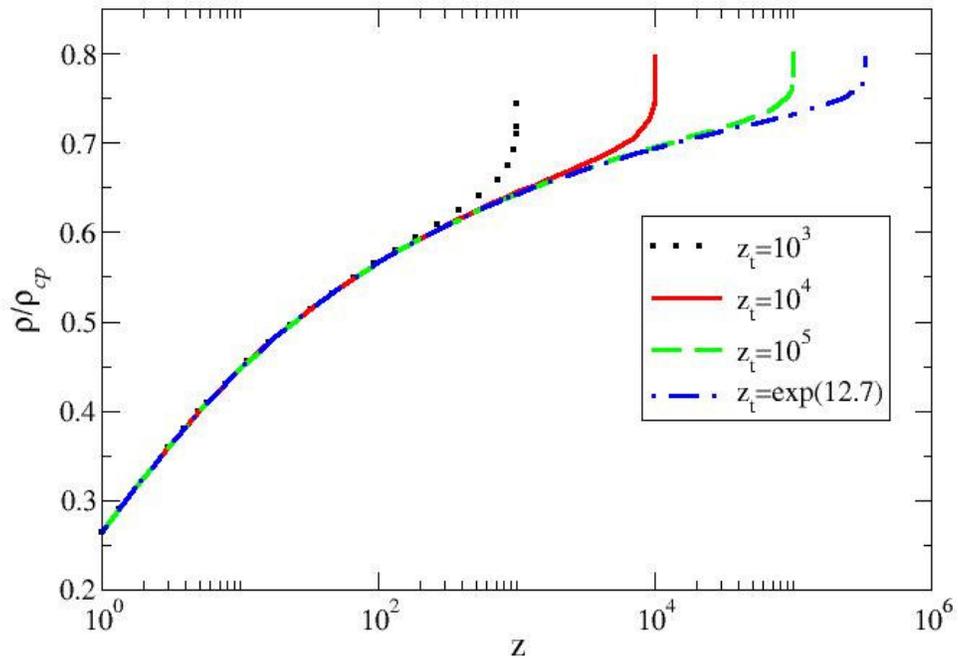

**Figure 3**: The equation of state for the hard-disks models, as calculated by our matrix method (line). The data agrees well with Monte-Carlo results (symbols). The insets presents the critical behavior at the vicinity of the termination point: the results of the matrix method (symbols) and the fit to a power law $p_c - p \approx (\rho_c - \rho)^{1/(1-\alpha)}$, with $1/(1-\alpha) = 4.55$, or $\alpha = 0.78$.

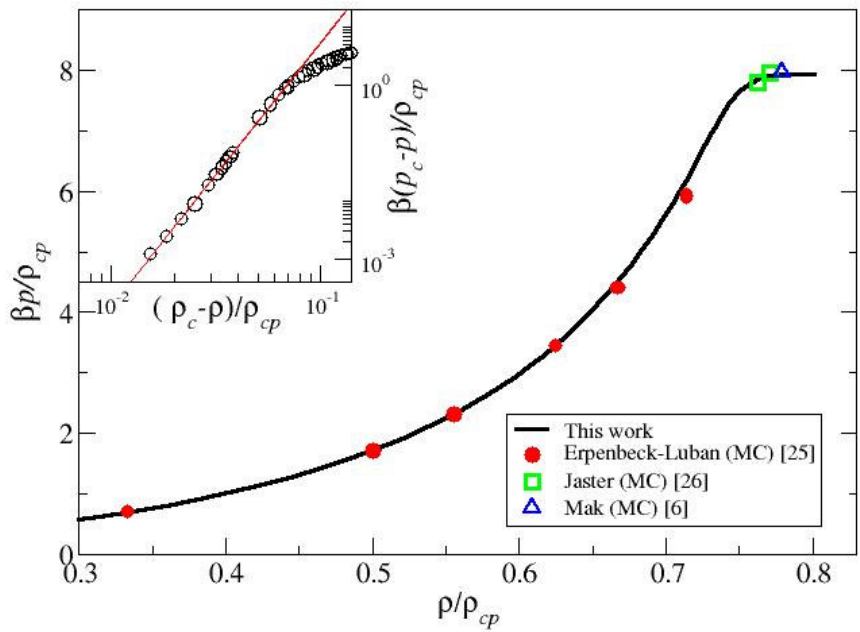